\documentclass{ws-procs9x6}            
\usepackage[utf8]{inputenc}
\usepackage{multirow}
\newcommand{\RomanNumeralCaps}[1]
    {\MakeUppercase{\romannumeral #1}}
\begin{document}
\title{Lifetime and dominant decay modes\\
of the tetraquark with double beauty $bb\bar u\bar d$
\footnote{%
To appear in the 
Proceedings of the \RomanNumeralCaps{18} International Conference on Hadron Spectroscopy 
and Structure (HADRON2019),  Guilin, China, 16-21 Aug. 2019. 
Talk given by J.-M.~Richard}
}

\author{J.-M.~Richard$^\dagger$}

\address{Universit\'e de Lyon, Institut de Physique des 2 Infinis de Lyon, IN2P3-CNRS--UCBL,\\ 
4 rue Enrico Fermi, 69622  Villeurbanne, France\\
$^\dagger$E-mail: j-m.richard@ipnl.in2p3.fr}

\author{E.~Hern\'andez and A.~Valcarce}

\address{Departamento de F\'\i sica Fundamental e IUFFyM, Universidad de Salamanca, \\
37008 Salamanca, Spain}

\author{J.~Vijande}

\address{Unidad Mixta de Investigaci\'on en Radiof{\'\i}sica e Instrumentaci\'on Nuclear 
en Medicina\\ (IRIMED),  Instituto de Investigaci{\'o}n Sanitaria La Fe (IIS-La Fe)-Universitat de Valencia (UV) 
and IFIC (UV-CSIC) 46100 Valencia, Spain}

\begin{abstract}
We briefly review the stability of the $QQ\bar q\bar q$-type of tetraquarks with two heavy quarks and 
two light antiquarks. We present the first comprehensive estimate of the lifetime and leading decay 
modes of the exotic meson $bb\bar u\bar d$ with double beauty.
\end{abstract}


\bodymatter

\section{Introduction}\label{sec:intro}
There is an  exciting activity dealing with exotic or crypto-exotic hadrons containing heavy 
quarks and/or antiquarks. Most experimental searches and results deal with hidden-flavor states 
such as the $XYZ$  mesons or $P_c$ baryons, which in some cases are remarkably narrow, but still 
unstable with respect to dissociation into lighter hadrons. 
There is also a renewed interest in flavor-exotic mesons which were already foreseen decades ago and 
seemingly will become accessible to the next generation of experiments. It is predicted that 
some $QQ\bar q\bar q$ states with two heavy quarks and two light antiquarks lie below their 
lowest dissociation threshold, and thus will provide the first evidence for \emph{stable} multiquark 
states beyond the familiar nuclei. Some relevant reviews 
include Refs.~\citenum{Chen:2016qju,Lebed:2016hpi,Richard:2016eis,Ali:2019roi,Brambilla:2019esw,Tanabashi:2018oca}.

The physics of weak interactions will be much stimulated by the occurrence of such states. The decay of 
strangeness has revealed the possibility of oscillations, the Cabibbo mixing scheme, and the dominance 
of nonleptonic modes for hyperon decays, due to the strong final state interaction. The decay of 
charmed hadrons has shown how the lifetimes and the semileptonic branching ratios are sensitive to 
the light constituents surrounding the charmed quark, with very different values for $D^+$ and $D^0$, 
and for the various charmed baryons. In contrast, the hadrons with single beauty, such as the $B$ mesons 
and the $\Lambda_b$ baryons have a remarkably similar lifetime $\tau\sim 1.5\,$ps, suggesting that the $b$ quark, 
while decaying, ignores its environment.

Our aim is to review briefly the stability of the $T_{bb}^- \equiv bb\bar u\bar d$ isoscalar exotic meson in constituent 
models, and to study its decay modes. Naively, a hadron with two $b$ quarks should decay about 
twice faster than a hadron with a single $b$. This is the result found for $bbq$ baryons~\cite{Kiselev:2001fw}.  
In the case of the tetraquark, two effects might act in the opposite direction. Firstly, for a typical decay 
$T_{bb}\to \bar B^{(*)}+D^{(*)}+X$, a fraction of the released energy is taken by the spectator $\bar B^{(*)}$, 
and less effective phase-space is left for the weak decay itself. It is known, from many examples in the study 
of nuclear $\beta$ decays, how sensitive are the weak interactions upon the available phase space. Secondly, 
after the $b\to c$ transition, the hadronization is somewhat more laborious than in the case of $B$ decay, 
because the antiquarks are further apart and not always in an optimal color state. This is why a detailed 
calculation is necessary, using realistic wave functions for both the tetraquark and the hadrons in the 
final state.\cite{Hernandez:2019eox} 
\section{Stability of doubly-heavy tetraquarks}
The first theoretical indication for the stability of $QQ\bar q\bar q $ tetraquarks was made nearly four decades 
ago~\cite{Ader:1981db}, and confirmed by several groups, as reviewed, e.g., in Ref.~\citenum{Richard:2016eis}. 
It can be understood in the following way. Let us denote by $M$ and $m$ the constituent masses, and start from the 
case where $M=m$, being first restricted to a spin-independent interaction, which is assumed to be \emph{flavor 
independent}. If $M$ is increased and $m$ decreased with the average inverse mass kept constant, the energy of 
the tetraquark is lowered by the breaking of symmetry (namely charge conjugation), while the threshold energy 
remains constant. For a large enough mass ratio $M/m$, stability is reached. This is the same mechanism by 
which the hydrogen molecule is much more stable than the positronium molecule, when the bindings are scaled 
by their respective threshold energies.  

Explicit four-body calculations show that this chromoelectric mechanism, though fully rigorous, it is not 
sufficient to provide binding, even for $bb\bar q\bar q$. Fortunately, there is another effect, due to 
chromomagnetic forces. In an isoscalar $QQ\bar u\bar d$ state with $J^P=1^+$, the $\bar u\bar d $ pair is 
mainly in a spin $s_{ud}=0$ and orbital momentum $\ell_{ud}=0$ state, and benefits from a strong spin-spin 
attraction. In the threshold, the spin-spin effects are penalized by the mass of the heavy quark. 

Note that we do not have to \emph{assume} any diquark clustering beforehand. We simply note that the dynamics 
induces some correlation between the two heavy quarks, due to chromoelectricity, and between the two light 
antiquarks, due to chromomagnetism. 
Pairwise interactions based on color-octet exchange
induce mixing between the $\bar33$ and $6\bar6$ states in the $QQ-\bar q\bar q$ basis,
enhancing the $\bar33$ components for larger values of $M$ due to the
attractive chromoelectric interaction of the $QQ$ pair that it is absent
in the $Q\bar q$ threshold. This result is only valid in the bottom sector. 
In the charm sector, the binding mechanism is different: the $\bar33$
and $6\bar6$ components have a similar probability and are mixed by the 
chromomagnetic interaction. 
A brute-force diquark approximation might be misleading.\cite{Richard:2018yrm}

Within explicit quark models, the state of art of rigorous calculations is the following. The double-charm isoscalar
tetraquark $cc\bar u\bar d$ with $J^P=1^+$ is probably slightly bound below its formal threshold $DD^*$~\cite{Rosina:2004wqe}, 
i.e., should be searched for as a peak in $DD\pi$ or $DD\gamma$. The double-beauty isoscalar $T_{bb}$ is stable,
lying approximately 150\,MeV below the strong decay threshold $B^-\bar {B^*}^{0}$ and 105\,MeV below 
the electromagnetic decay threshold $B^- \bar B^0 \gamma$. 
Thus it decays weakly.\cite{Hernandez:2019eox}\@
The isoscalar $bc\bar u\bar d$ might be stable too, and if this is the case, can bear 
$J^P=0^+$ or $1^+$. While the $J^P=0^+$ state would be strong and electromagnetic-interaction stable, the $J^P=1^+$
would decay electromagnetically to $\bar B D \gamma$.\cite{Carames:2019,Francis:2019}
\section{$T_{bb}$ weak decays}
\begin{figure}[ht!]
\vskip -.3cm
 \centerline{
 \includegraphics[width=.49\textwidth]{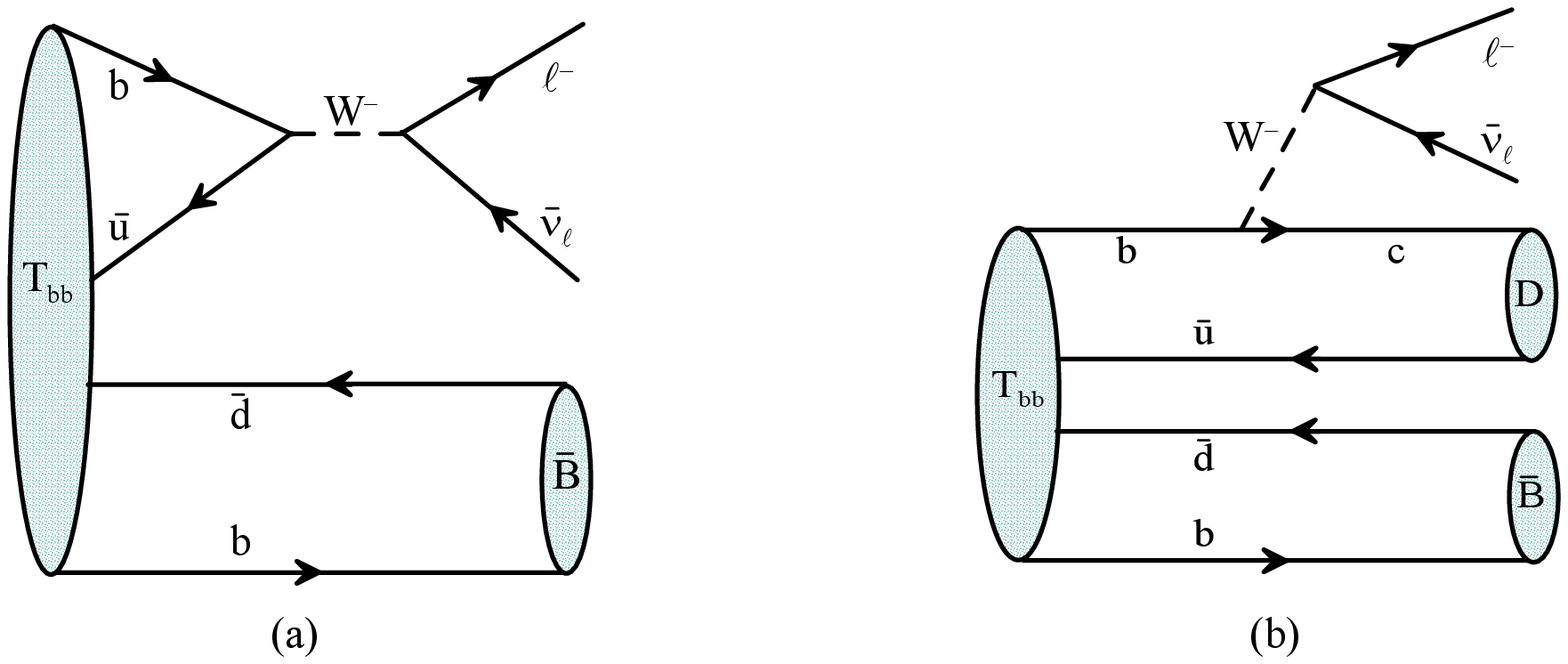}
 \qquad
\includegraphics[width=.49\textwidth]{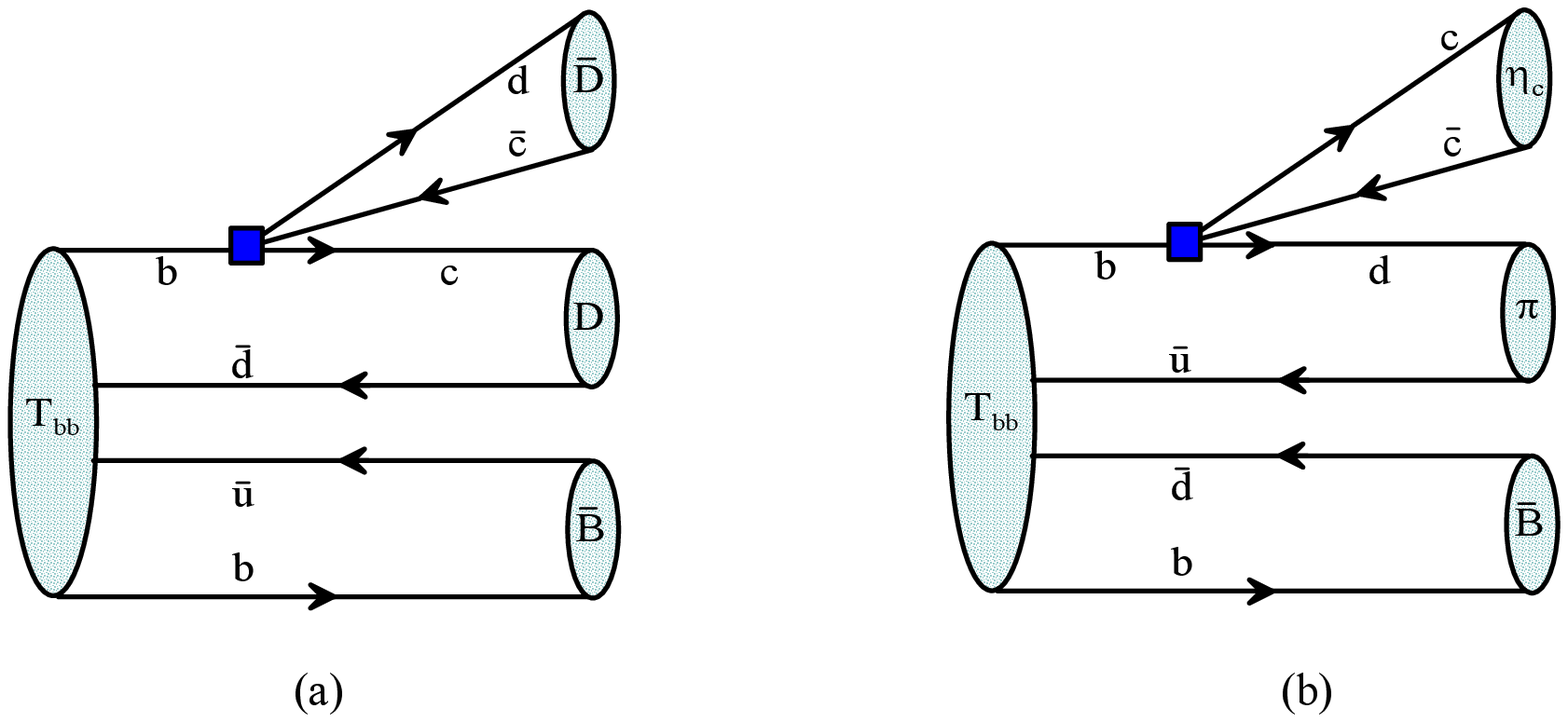}}
 \caption{Representative diagrams for semileptonic (left) and nonleptonic (right) decays of the $T_{bb}$ tetraquark}
 \label{fig1}
\end{figure}

We have calculated the width of the isoscalar $T_{bb}$ tetraquark due to all plausible semileptonic 
and nonleptonic decay modes.\cite{Hernandez:2019eox}\@ The hadronic decays are calculated within the factorization approximation. Some of the corresponding processes are illustrated in Fig.~\ref{fig1}.
The largest partial widths are found to be of the order of $10^{-15}$ to $10^{-14}\,$GeV. For the 
semileptonic modes, the corresponding decays are of the type 
$\bar B{}^{*} D^{*}\,\ell^-\bar\nu_\ell$, where $\ell=e$ or $\mu$.
Due to the large phase space available in all cases, the differences 
among the widths into the three lepton families are very small.

For semileptonic decays involving two mesons in the final state, the processes 
involving a $ b \to c$ vertex are favored compared to those involving a $b \to u$ vertex, 
due to the larger CKM matrix element. For the former, see table \ref{tab1}, and, 
for the latter, a typical estimate $\Gamma(T_{bb}\to \bar {B^*}^{0} \, \ell^- \, \bar \nu_\ell )\sim 0.04\times 10^{-15}\,$GeV.      
Besides, due to spin recoupling coefficients, the largest decay widths appear for vector 
mesons in the final state. 

\begin{table}
	\tbl{Decay widths, in units of $10^{-15}$\,GeV, of the leading semileptonic modes of $T_{bb}$}
 {\begin{tabular}{lclc}\hline
	Final state &$\Gamma$ &Final state &$\Gamma$\\
	\hline
	${B^*}^{-} \, {D^*}^{+} \, \ell^- \, \bar \nu_\ell $ & \multirow{2}{*}{$9{.}02 \pm 0{.}07 $}&
	${B^*}^{-} \, {D^*}^{+} \, \tau^- \, \bar \nu_\tau $&\multirow{2}{*}{$1.55\pm0.01 $}\\
	$\bar {B^*}^{0} \, {D^*}^{0} \, \ell^- \, \bar \nu_\ell $ && $\bar {B^*}^{0} \, {D^*}^{0} \, \tau^- \, \bar \nu_\tau $\\
	${B^*}^{-} \, D^{+} \, \ell^- \, \bar \nu_\ell $   & \multirow{2}{*}{$3{.}59 \pm 0{.}03$}&
	${B^*}^{-} \, D^{+} \, \tau^- \, \bar \nu_\tau$     &\multirow{2}{*}{$ 0.727\pm0.005$}\\
	$\bar {B^*}^{0} \, D^{0} \, \ell^- \, \bar \nu_\ell $&&  $\bar {B^*}^{0} \, D^{0} \, \tau^- \, \bar \nu_\tau $ \\
	$B^{-} \, {D^*}^{+} \, \ell^- \, \bar \nu_\ell $ &\multirow{2}{*}{$4{.}63 \pm 0{.}05$}&
	$B^{-} \, {D^*}^{+} \, \tau^- \, \bar \nu_\tau $&\multirow{2}{*}{$ 0.86\pm0.007$} \\
	$\bar B^{0} \, {D^*}^{0} \, \ell^- \, \bar \nu_\ell $&&$\bar B^{0} \, {D^*}^{0} \, \tau^- \, \bar \nu_\tau $\\
	$B^{-} \,  D^{+} \, l^- \, \bar \nu_l $& \multirow{2}{*}{$1{.}92 \pm 0{.}02$}&
	$B^{-} \,  D^{+} \, \tau^- \, \bar \nu_\tau $&\multirow{2}{*}{$ 0.409\pm0.003$}\\
	$\bar B^{0} \, D^{0} \,  \ell^- \, \bar \nu_\ell $&&$\bar B^{0} \, D^{0} \,  \tau^- \, \bar \nu_\tau $\\ \hline
\end{tabular}}
	\label{tab1}
\end{table}

For the nonleptonic decays, the largest widths are for the decays of 
the type $\bar B{}^{*} D^{*} D_s{}^{*}$. 
All of them contain a $b \to c$ vertex and a $D^{*}_{s}$ meson in the final
state. See Table~\ref{tab2}. Once again  vector mesons are favored in the final state. 
Processes with $D_s$ or a light meson final states arising from vacuum have 
decay widths comparable to the corresponding semileptonic decay.
\begin{table}
	\tbl{Decay widths, in units of $10^{-15}$\,GeV, of the leading nonleptonic modes of $T_{bb}$}
{\begin{tabular}{lclc}
 	\hline
	Final state &$\Gamma$ &Final state &$\Gamma$\\
	\hline
${B^*}^{-} \, {D^*}^{+} \, D_s^-$       & \multirow{2}{*}{$4{.}00 \pm 0{.}06$} &
${B^-} \, {D^*}^{+} \, {D_s^*}^- $      & \multirow{2}{*}{$3{.}15 \pm 0{.}05$}\\
$\bar {B^*}^{0} \, {D^*}^{0} \, D_s^- $                         && $\bar {B^0} \, {D^*}^0 \, {D_s^*}^- $  \\  
${B^*}^{-} \, {D^*}^{+} \, {D_s^*}^-$  & \multirow{2}{*}{$6{.}50 \pm 0{.}09$} &
$B^- \, D^+ \, {D_s^*}^- $    & \multirow{2}{*}{$1{.}20 \pm 0{.}02$}\\
$\bar {B^*}^{0} \, {D^*}^{0} \, {D_s^*}^- $ &&          $\bar {B^0} \, D^0 \, {D_s^*}^- $\\
${B^*}^{-} \, D^{+} \, D_s^-$& \multirow{2}{*}{$2{.}57 \pm 0{.}04$} &${B^*}^{-} \, {D^*}^{+} \, \rho^-$  & $3{.}57 \pm 0{.}09$ \\
$\bar {B^*}^{0} \, D^0 \, D_s^- $&& ${B^*}^{-} \, {D^*}^{+} \, \pi^-$& $1{.}28 \pm 0{.}03$ \\
${B^*}^{-} \, D^{+} \, {D_s^*}^- $ & \multirow{2}{*}{$2{.}32 \pm 0{.}03$} &
       ${B^*}^{-} \, D^+ \, \rho^-$                                    & $1{.}70 \pm 0{.}04$ \\
$\bar {B^*}^{0} \, D^0 \, {D_s^*}^- $ &&${B^*}^{-} \, D^+ \, \pi^-$& $0{.}70 \pm 0{.}02$ \\
$B^- \, {D^*}^{+} \, D_s^-$& \multirow{2}{*}{$2{.}78 \pm 0{.}05$} &$B^- \, {D^*}^{+} \, \rho^-$& $2{.}01 \pm 0{.}05$ \\
$\bar {B^0} \, {D^*}^{0} \, D_s^- $&& $B^- \, {D^*}^{+} \, \pi^-$& $0{.}77 \pm 0{.}03$ \\
\hline
\end{tabular}}
\label{tab2}
\end{table}

In the event where an isoscalar $T_{bc}$ tetraquark with $J^P=0^+$ exists, then 
the decay $T_{bb} (1^+) \to T_{bc}(0^+)\ell^-\nu$ is about $3.\times 10^{-15}\,$GeV.
The semileptonic decay to the isoscalar $J^P=0^+$ tetraquark $T_{bc}^0$ is relevant but
it is not found to be dominant.

We also estimated all plausible decay modes such as $\bar B{}^0 e^-\bar\nu_e$, or $B{}^{*-}D^+\pi^-$, etc. 
The total width turns out to be about $87\times 10^{-15}\,$GeV, which gives an 
upper bound for the lifetime of about $7.6\,$ps.
\section{Outlook}

We have examined the stability of $QQ\bar q \bar q$ tetraquarks, showing how the structure of 
the $T_{QQ}$ state evolves from a molecular-like system to a compact-like structure when 
moving from the charm to the bottom sector, due to a competition between chromoelectric and 
chromomagnetic forces.
We have also examined a variety of decay modes of $T_{bb}$, either semileptonic and nonleptonic. The leading modes 
involve a $\bar B{}^{(*)}$ and a $D ^{(*)}$ in the final state.  The experimental search might look at channels 
with a smaller branching ratio but an easier identification, as the final states pointed out in Refs.~\citenum{Xing:2018bqt,Ali:2018xfq,Ali:2018ifm}.  

The most striking result is our estimate of the lifetime
\begin{equation}
 \tau(T_{bb})\sim 7.6\,\mathrm{ps}~,
\end{equation}
to be compared with about $1.5\,$ps for hadrons with single beauty. 
For instance, if one searches for double-$b$ hadrons by the method of displaced  
vertex~\cite{Gershon:2018gda}, the tetraquark will  give a signature rather 
different from that of the baryons. See, also, the discussion in Refs.~\citenum{Ali:2018xfq,Ali:2018ifm}.


\end{document}